\documentclass[a4,prd,amssymb,twocolumn,floatfix]{revtex4}
\usepackage{graphicx}

\begin{document}
\preprint{CERN-PH-TH/2007-158}
\title{Compact Stars as Dark Matter Probes}
\author{Gianfranco Bertone}
\affiliation{Institut d'Astrophysique de Paris, UMR 7095-CNRS,
Universit\'e Pierre et Marie Curie, 98bis boulevard Arago, 75014 Paris, France}
\email{bertone@iap.fr}
\author{Malcolm Fairbairn}
\affiliation{TH Division, Physics Department, CERN, Geneva, Switzerland}
\email{malc@cern.ch}

\begin{abstract}
We discuss the consequences of the accretion of dark matter (DM) particles on compact
stars such as white dwarfs and neutron stars. We show that in large regions of
the DM parameter space, these objects are sensitive probes of the presence of DM,
and can be used to set constraints both on the DM density and on the physical
properties of DM particles.
\end{abstract}

\maketitle

\section{Introduction}

Observations continue to provide overwhelming evidence for a {\it dark} component of matter in the Universe~\cite{Bertone:2004pz,Bergstrom:2000pn}.  There are a number of different candidates for this dark matter (DM), for example, weakly interacting massive particles (WIMPs) are attractive candidates because their weak scale annihilation cross-section naturally gives rise to densities in today's universe comparable with what is needed cosmologically.  They are also interesting candidates in as much as they may be probed by the increasingly accurate direct detection experiments in development such as the current leader XENON10 \cite{xenon10}.  Furthermore, it is possible that WIMPs could be produced at the soon to be operating Large Hadron Collider (LHC) at CERN. 

There also exist other candidates which are more weakly coupled to standard model particles.  The axion is an example of a very feebly coupled cold DM candidate, whereas the gravitino or the sterile neutrino can be candidates for either cold or warm DM, depending on their production mechanisms and masses and are equally difficult to detect (see Refs. ~\cite{Bertone:2004pz,Bergstrom:2000pn} and references therein).  Another set of DM particles which have been suggested are superheavy particles known as WIMPzillas which are produced non-thermally at the end of inflation \cite{igor,chung}.

It has long been realised that a finite DM-nucleon cross section would result in scattering between DM and the gas in stars, knocking the DM particles into orbits which result in their subsequent capture \cite{steigman,spergel,gould,griest}.  Such captured particles would "sink" to the centre of stars where they would annihilate with themselves. Indeed neutrinos from such annihilations at the centre of the sun will be searched for at neutrino telescopes such as ICECUBE, currently under construction in Antarctica \cite{bergstrom96,landsman}.

The amount of DM in the solar system is such that while one can expect these products of WIMP annihilation in the centre of the Sun, the total injected energy rate associated with this annihilation is a tiny fraction of the luminosity in the sun due to nuclear burning.  We do not therefore expect that the presence of these WIMPs will affect the way that the Sun itself burns, although future helioseismic observations may probe an interesting region of the WIMP parameter space\cite{Lopes:2002gp}.  Here, we focus on situations where the capture of WIMPs may significantly affect the strcture and/or the evolution of the astronomical objects onto which they are captured.

N-body simulations of DM halos suggest that the DM density close to the galactic centre should be larger by several orders of magnitude than the density in the solar system (see e.g. the recent "Via Lactea" simulation of Diemand et al. \cite{Diemand:2006ik} and references therein).  Stars in those regions therefore may experience the accretion of a large amount of DM which could change the internal energy exchange mechanisms, e.g. changing the temperature gradient in convective stellar cores \cite{salati}.  Furthermore, if the density of DM is extremely high, the energy injection rate due to DM annihilations could compete with the usual nuclear energy production rate \cite{moskwai1,moskwai2}. As for the Earth, the requirement that the injected energy does not exceed the measured Earth's heat flow, has been used to set interesting constraints on the scattering cross-section of DM particles off nuclei~\cite{mack}.

The capture of DM onto stars is proportional to the number of nucleons in the star times the escape velocity, therefore compact objects such as white dwarfs or neutron stars are ideal targets for searches aimed at detecting the effects of DM accretion.
Both classes of degenerate compact objects are unfortunately also usually rather hot, so that any increase in temperature due to the accretion of WIMPs would be difficult to detect.   
In the past decade, however, great advances have been made in the detection of cool white dwarfs in globular clusters, where very old stars with temperatures of less than $4000$ K have been observed~\cite{richerm4}.  It would seem therefore to be a good place to look for the effects of WIMP accretion and indeed we will show that the constraints which can be obtained are very interesting, despite being model dependent.

As for neutron stars, we will see that it is rather difficult to detect the heating produced by the annihilation of WIMPs captured in their cores, since the total cross section for capture is limited by the relatively small size of the star.  The accretion of DM particles which are not their own anti-particles or do not annihilate with themselves very quickly is however constrained, since in this case DM particles (if they are more massive than nucleons and have couplings to them) would accumlate at the centre of the neutron star, eventually becoming the dominant source of gravitational potential in that central region.  As we shall see, if the density of DM in the core were to increase more rapidly than its self annihilation rate, then the accumulation could lead to a gravitational instability and the collapse of the DM cloud.

The paper is organised as follows: in the next section we will discuss the accretion of DM particles onto compact objects in generality. In section \ref{wd} we will then develop a model for the baryons and the DM in globular clusters, specifically M4, where observations of cool stars have been made close to the centre of the cluster where the DM, if there at all, is expected to be densest.  We argue that if there is DM in globular clusters, then its density at the radius we are interested in is an extremely weak function of the total mass within the cluster.  We work out the accretion rate of DM onto white dwarfs and estimate how it would change their luminosity, finally comparing with the actual data to place constraints on the combination of DM density and WIMP-nucleon cross section.

Then in section \ref{ns} we will describe the various possible effects upon neutron stars due to the accumulation of DM.  We calculate the expected luminosity of galactic neutron stars at different distances from the centre of the milky way, before considering the interesting effects which could happen inside neutron stars due to the accretion of a DM species with a small self annihilation cross section.

\section{Accretion of DM onto gravitating objects}

The accretion of WIMPs onto stars has been studied for many years now \cite{steigman,spergel,gould,griest}. The capture rate $Gamma_c$ which we adopt interpolates between the two 
different capture rates defined for optically thin and thick bodies in 
Ref.~\cite{bottino}
\begin{equation}
\Gamma_c=\left(\frac{8}{3\pi}\right)^{1/2}\frac{\rho_{dm}\bar{v}}{m_{dm}}\left(\frac{3v_{esc}^2}{2\bar{v}^2}\right)\sigma_{eff}
\label{capture}
\end{equation}
where $\rho_{dm}$ is the density of DM round the star, $m_{dm}$ is the mass of the DM particles and $\bar{v}$ is the average DM velocity.  $v_{esc}$ is the escape velocity of the astronomical body in question.

The effective cross section $\sigma_{eff}$ is given by by one of two things, normally it is the sum of the cross sections upon the individual nuclei in the star, including the coherence factor which means that the spin independent cross section of a nucleus of element $i$ with $A_i$ nucleons will be $A_i^4$ times the cross section $\sigma_{si}$ of an individual proton or neutron (see e.g. \cite{giudice,lewin}).  However, this total cross section cannot be larger than the geometrical size of the star, so the effective cross section is given by whichever is smallest
\begin{equation}
\sigma_{eff}=min\left[\sigma_{si}\sum_i\frac{M_*}{m_p}\frac{x_i}{A_i}A_i^4\quad,\quad\pi R_*^2\right]
\end{equation}
where $x_i$ is the mass fraction of element $i$, $M_*$ and $R_*$ are the mass and radii of the star and $m_p$ is the proton mass. For the capture of WIMPs onto neutron stars, one expects that the geometrical cross section may be less than the sum of the individual cross sections.  The cross section for scattering between the DM particle and nucleons would have to be less than $10^{-45}$ cm$^2$ in order for the sum of the individual cross sections in the star to be less than the geometrical cross section of the star.

Once DM is captured by a star or a compact object, it will form a thermal distribution inside the star of characteristic radius \cite{griest}
\begin{equation}
r_{th}\sim\left(\frac{9kT_c}{4 \pi G \rho_c m_{dm}}\right)^{1/2}
\end{equation}
where $\rho_c$ and $T_c$ is the central density and temperature and $m_{dm}$ is the mass of the DM particle.  Typical values of this thermal radius for WIMP DM will be around a metre for a neutron star or a kilometre for a cool white dwarf.  Having been concentrated in the centre of the star, the DM will annihilate with itself, the equation for the evolution of number density in the star over time is
\begin{equation}
\frac{dN}{dt}=\Gamma_c-\Gamma_a
\label{evol}
\end{equation}
where the annihilation rate $\Gamma_a$ is given by
\begin{equation}
\Gamma_a=\frac{1}{2}\frac{N^2(\sigma_{ann} v )}{\frac{4}{3}\pi r_{th}^3}
\end{equation}
which contains the DM self annihilation cross section $(\sigma_{ann} v)$ and the total number of DM particles in the star $N$.  When $\Gamma_c=\Gamma_a$ the capture and annihilation rates will be equal and any DM which accretes onto the star will be instantly converted into additional luminosity. The timescale for this steady state to be reached is given by
\begin{equation}
\tau_{eq}=\frac{N}{\sqrt{\Gamma_a\Gamma_c}}.
\end{equation}
In the next section we will go on to calculate the effect of DM accretion onto cool white dwarfs.

\section{White dwarfs in globular clusters\label{wd}}

The presence of DM in globular clusters is a controversial issue, with early work by Peebles \cite{peebles} suggesting that globular clusters form along with dark halos.  Since then there are those that argue that even if the clusters did start with a halo of DM that it will have been tidally disrupted since then due to interactions with the host galaxy \cite{saitoh}.  However, other studies suggest that the cores of DM halos in globular clusters will survive successive tidal interactions with the host galaxy \cite{sills}.  The combination of weakly interacting DM and old cool white dwarfs therefore seems to present a real possibility of testing the properties of DM.

The distribution of stars in clusters can be fit well by solutions of the collisionless Liouville equation for a given velocity distribution \cite{king3}.  The cluster profiles are then determined by the core radius $r_c$ and the tidal radius $r_t$ plus an overall normalisation. The relevant parameters for the globular cluster M4 are a core radius of $r_c=0.83'$ in arc-minutes and a concentration parameter $c=\log(r_t/r_c)=1.59$ \cite{harris}.  This, combined with the distance to the cluster of $1.73$ kpc \cite{richerm4}, gives $r_c=0.417$ pc.  We use these parameters to model the density of stars in the globular cluster using the King model. The overall normalisation of density is obtained by integrating the King profile and setting the total mass to $10^5M_\odot$, this figure obtained from the apparent magnitude and the distance which give a total luminosity of the cluster of $L\sim1.5\times 10^5L_\odot$.

In order to estimate the size and the extent of the DM halo surrounding the globular cluster, we follow the procedure of \cite{sills}. 
The average cosmological ratio between baryons and DM is $\Omega_b/\Omega_{DM}\sim0.2$ \cite{WMAP}, while the fraction of baryons in the form of globular clusters is $F = 0.0025$ \cite{globprob}.  Mashchenko and Sills have therefore estimated that, having taken into account mass loss during stellar evolution (and other factors, see Ref. \cite{globprob} for details), the ratio of baryonic to non-baryonic mass is $F=0.0088$.  Here, we assume that the mass of the DM in the Globular cluster is $M_{DM}=10^7 M_\odot$, and we will show below that our results are rather insensitive to the exact value of the baryonic fraction.

\begin{figure}
\begin{center}
\includegraphics[width=0.65\textwidth]{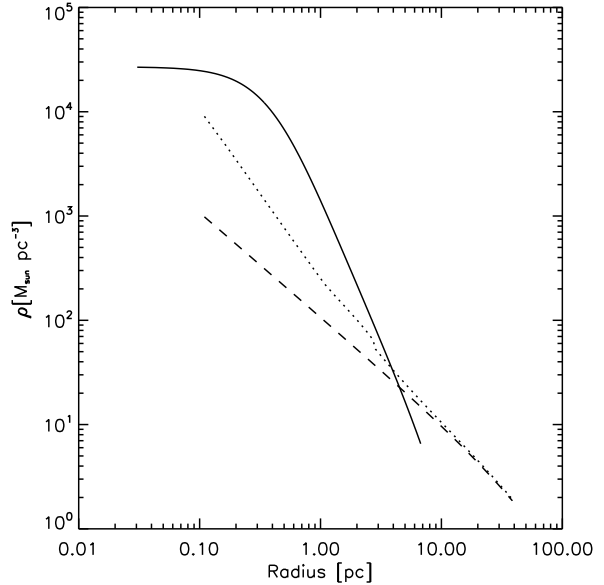} 
\caption{Density of baryons and DM in the central region of the globular cluster M4.  The solid line shows the density of baryons as a function of radius.  The dotted and dashed lines represent the density of DM for the normal NFW profile and for an adiabatically contracted profile, respectively.}
\label{density}
\end{center}
\end{figure}

We define the virial radius $R_{vir}$ as being the radius within which the average density is bigger than the average cosmological density by a factor $\Delta$, called {\it virial overdensity}, for which we have adopted here the fitting form of Bryan and Norman\cite{bryannorman}
\begin{equation}
\Delta= 18\pi^2+82(\Omega_m-1)-39(\Omega_m-1)^2\simeq101 \, ,
\end{equation}
leading to the relationship
\begin{equation}
R_{vir}=\left(\frac{3M_{DM}}{4\pi\delta\rho_{crit}}\right)^{1/3} \simeq 5.5 \, \rm kpc \, ,
\end{equation}
where we have assumed a Hubble constant of 72 km s$^{-1}$Mpc$^{-1}$ \cite{hst}, and $\Omega_m=0.3$.  A virial radius at $z=0$ of more than 5 kpc does not really make any sense because globular clusters are well embedded within the much larger halo of the milky way, but evidence from N-body simulations suggests that the inner part of DM halos survives tidal stripping \cite{sills}, and we will only use $R_{vir}$ to derive the properties of the initial central DM distribution.  Following \cite{bullock,ullio}, we will assume a concentration parameter for a halo of this mass to be $c=32$ (slightly different models give rise to different $c$ values, but such changes will not change the results significantly) and we will assume a Navarro, Frenk and White profile of the form \cite{nfw}
\begin{equation}
\rho(r)=\frac{\rho_c}{\frac{r}{a}\left(1+\frac{r}{a}\right)^2}
\end{equation}
with $a=171$ pc and 
\begin{equation}
\rho_c=\frac{M_{DM}}{4\pi a^3}\left[\ln(1+c)-\frac{c}{1+c}\right]^{-1} \simeq 0.63 M_\odot \rm{pc}^{-3}
\end{equation}
which finishes our definition of the DM profile. The different density components are shown in Fig. \ref{density}.

We will be interested in the density of DM close to the core radius $r_c$ of the globular cluster.  As we claimed earlier, this is a very weak function of the total mass of the halo, $M_{DM}$.  In fact $\rho(r_c)$ changes by around a factor 3 for different halo masses between $10^6M_\odot$ and $10^8M_\odot$.  The reason for this weak dependence on the total mass is that the density of DM at some small fixed radius $r\ll a$ is proportional to $\rho_c a$ which in turn is proportional to $M_{DM}^{1/3}c^2/(\ln(1+c)-c/(1+c))$ and as $M_{DM}$ goes down, the halo profile concentration $c$ increases, see for example \cite{ullio}.  This will make our results less sensitive to the overall mass of DM.

The DM halo will be distorted by the presence of baryons, leading to an enhancement of the halo profile in the inner regions of the globular cluster 
\cite{blumenthal,gnedin,jesper}.  At the same time, the heating of DM particles due to interactions with stars will tend to wipe out the effect of adiabatic contraction, over a timescale \cite{merritt}
\begin{equation}
T_{heat}\equiv\left|\frac{1}{\epsilon}\frac{d\epsilon}{dt}\right|^{-1}=\frac{0.814 v_{rms}^3}{G^2m_*\rho_*\ln\Lambda}
\end{equation}
where $\epsilon$ in the kinetic energy of a typical DM particle, $\rho_*$ and $m_*$ are the density and mass of stars, $G$ is Newton's constant, $v_{rms}$ is the rms velocity of the DM particles which we assume is equal to that of the stars and $\Lambda\sim 0.4 N$ where $N$ is the total number of stars below the radius in question.  
At the core radius of the cluster, the timescale for heating of DM is much less than the age of the universe, so one cannot expect to rely upon the density profiles plotted in figure \ref{density} since globular cluster $M4$ is nearly as old as the Universe.  We will therefore assume a maximum density of 50 $M_\odot$ pc$^{-3}$, which is the approximate density at the radius where the timescale for heating of DM becomes larger than the age of the Universe.

A $0.5M_\odot$ white dwarf has a radius of approximately $10^9$ cm and will be composed of carbon or oxygen (the densities and observed temperatures of newly formed white dwarfs are inconsistent with the presence of hydrogen or helium, see e.g. \cite{tdlee}).  Since we wish to make as many conservative assumptions as possible in the calculation, and because the white dwarfs we are looking at are formed in low metallicity environments, we will assume that the white dwarfs in question are composed entirely from carbon.  We can then work out the spin independent cross section which would be required in order for the geometrical cross section to dominate over the sum of the individual nuclei cross section, which turns out to be around $10^{-41}$ cm$^2$.  Because the constraints on the spin independent WIMP-nucleon cross section from XENON are at the level of $10^{-43}$ cm$^2$, we will assume that $\sigma_{eff}$ is given by the sum of the individual nucleon cross sections rather than the geometrical cross section of the star.

Other ingredients that we need for our calculations are the escape velocity of the white dwarf, which is trivially obtained, and also the velocity of the DM particles at the radius corresponding to $r_c$.  Studies of the velocities of stars in globular cluster M4 show typically a velocity dispersion of a few kilometres per second \cite{peterson}.  Equation (\ref{capture}) shows us that the capture rate increases for low velocity DM, so to be conservative we choose the maximum velocity possible at $r_c$ which corresponds to the escape velocity obtained by integrating the derivative of the potential of the King profile of stars (as in the Milky Way, the inner potential of M4 is completely dominated by baryons).  This gives us a velocity for the DM of 20 km s$^{-1}$, assuming that there is no black hole at the centre of the cluster. The presence of such an object would anyway seem to be at odds with the observed velocity dispersion of stars.

The timescale for the equilibrium between the accretion and annihilation of WIMPs will be very roughly of the order of a year and the annihilation rate of WIMPs in the star will quickly reach a steady state such that equation (\ref{evol}) will be equal to zero and all of the DM incident on the star can be assumed to go instantaneously into increasing the luminosity of the star.  The luminosity of the star purely due to the accretion and subsequent annihilation of DM will therefore be given by ($3v_{esc}^2\gg 2\zeta \bar{v}^2$)
\begin{eqnarray}
L&=&\left(\frac{8}{3\pi}\right)^{1/2}\rho_{DM}\frac{3v_{esc}^2}{2\bar{v}}\sigma_{si}\sum_i\frac{M_*}{m_p}\frac{x_i}{A_i}A_i^4\nonumber\\
&=&3\times 10^{27} \left(\frac{\rho_{DM}}{50 M_\odot pc^{-3}}\right)\left(\frac{2M_*}{M_\odot}\right)^2\nonumber\\
&\times &\left(\frac{\sigma_{si}}{10^{-44}{\rm cm^2}}\right)\left(\frac{10^{9}{\rm cm}}{R_*}\right){\rm erg \, s^{-1}}
\end{eqnarray}
which can be regarded as the minimum expected luminosity of a star in an environment of weakly interacting, self annihilating DM of density $\rho_{DM}$ and velocity $\bar{v}$.

Observations of faint white dwarfs have been made in the central region of globular cluster M4 using the equivalent of UVI filters on the Wide Field/Planetary Camera (WFPC2) of the Hubble space telescope \cite{richerm4}.  Observations were taken in three fields, the closest one to the centre of the cluster with the planetary camera centered at $0.5r_c\sim 0.2$pc.

\begin{figure}
\begin{center}
\includegraphics[width=0.65\textwidth]{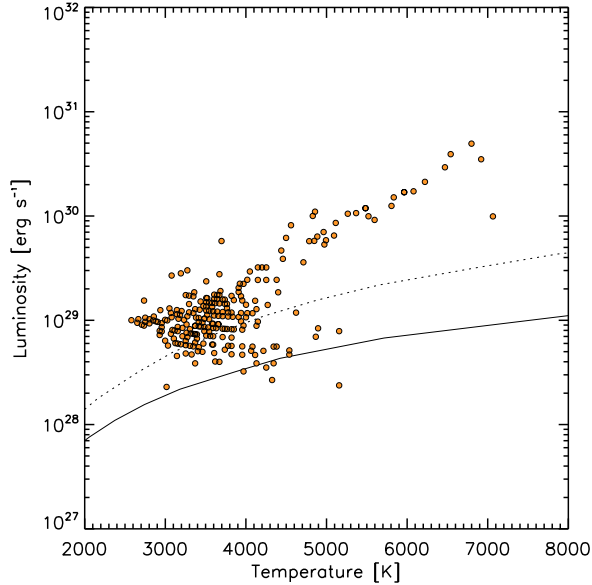} 
\caption{Luminosity vs. temperature for the white dwarfs in the inner field of the observations made in paper \cite{richerm4} (data points).  Also plotted (lines) is the minimum luminosity expected for white dwarfs radiating as black bodies, for WIMP-nucleon cross sections of $10^{-43}$ cm$^2$ and $10^{-44}$ cm$^2$.  See text for details.}
\label{wdplot}
\end{center}
\end{figure}

We assume that the White dwarfs are perfect black body emitters, and we use their V-I magnitudes to calculate their temperature. We then use this temperature, plus the distance to globular cluster M4, to calculate the absolute bolometric magnitude of the stars, and consequently their luminosity.  The resulting luminosities and temperatures of the white dwarfs are plotted in figure \ref{wdplot}. In the same field, there will be stars further away from the centre of the cluster than 0.2 pc.  However, due to the centrally peaked nature of the King profile, we calculate that more than 99\% of the stars along a line of site at angular distance from the centre of the cluster corresponding to 0.2 pc will be at radii less than 5 pc from the centre of the cluster and will therefore be surrounded by the same density of DM. 

We derive the radii of the white dwarfs using their temperatures and luminosities and the assumption that they (and the sun) are black-body emitters, i.e.
\begin{equation}
r=r_\odot\sqrt{\frac{L}{L_\odot}}\left(\frac{T_\odot}{T}\right)^2
\end{equation}
which fits well with the radius expected for $0.5M_\odot$ white dwarfs using the standard mass-radius relations for white dwarfs. 

In figure \ref{wdplot} we have also plotted two lines corresponding to the luminosity one would expect due to the accretion and subsequent annihilation of WIMPs if the white dwarfs were placed in a region where the DM density is 50 $M_\odot$ pc$^{-3}$, as we expect in the inner region of the globular cluster where they are located.  We plot lines rather than points because we calculate the expected luminosity and temperature for white dwarfs of different masses, and hence radii.  The two lines correspond to two different WIMP-nucleon cross sections, $10^{-44}$ cm$^2$ and $10^{-43}$ cm$^2$.  If such a density of WIMP DM did exist in the centre of globular clusters with a WIMP-nucleon cross section corresponding to one or other of the lines, then we would not expect to see any white dwarfs lying below the line.

Since the density of DM is robust to the total amount of matter in the cluster, this analysis suggests that if a core of DM remains in globular clusters then we are able to place a constraint upon the WIMP nucleon cross section between $10^{-43}$ and $10^{-44}$ cm$^2$ which is very competitive with the latest experiments.  An alternative, more conservative statement would be that it is possible with this analysis to constrain the combination of WIMP nucleon cross section and DM density in globular clusters.  It would therefore be interesting if WIMPs were detected in the laboratory with cross sections close to this level $10^{-44}$ cm$^2$, as it could open a new branch of DM astronomy using white dwarfs.

Since 1997, the same group who obtained the data used in this paper have made much more detailed studies of NGC 6397 but at a much larger radius from the centre of the cluster.  Their observations are rather better than those used in this paper, enabling them to observe the super cool white dwarfs becoming more blue as H2 forms in their atmosphere, blocking the emission of redder frequencies.  By obtaining colour-magnitude diagrams of this detail at different radii in the same cluster and comparing them we predict that it should be possible to significantly increase the sensitivity of these constraints.

\section{Capture of DM onto neutron stars\label{ns}}

In this section we will look at the accretion of DM particles onto compact objects and their subsequent effects.  As in the case of white dwarfs, the time scale for equilibrium between capture and annihilation will be very short compared to typical astrophysical timescales. 

The most obvious possible observable signal of the accretion of DM onto neutron stars is also the simplest to calculate, namely the heating of the star, with consequent increase of the surface temperature, due to the energy injected by annihilating DM particles.  We follow the same procedure as that outlined in the previous section for the case of white dwarfs.  For neutron stars and WIMP-nucleon cross sections close to the experimental limit, the surface area of the star is larger than the sum of the cross sections for the individual nuclei.  (Unlike a nucleus with a large atomic number, a medium with a constant number density of nucleons gives rise no resonant enhancement, although the presence of rod and sheet like structures in the neutron matter may give rise to an order of magnitude enhancement as they can for the neutrino-nucleon cross section \cite{horowitz}.)

\begin{figure}
\begin{center}
\includegraphics[width=0.65\textwidth]{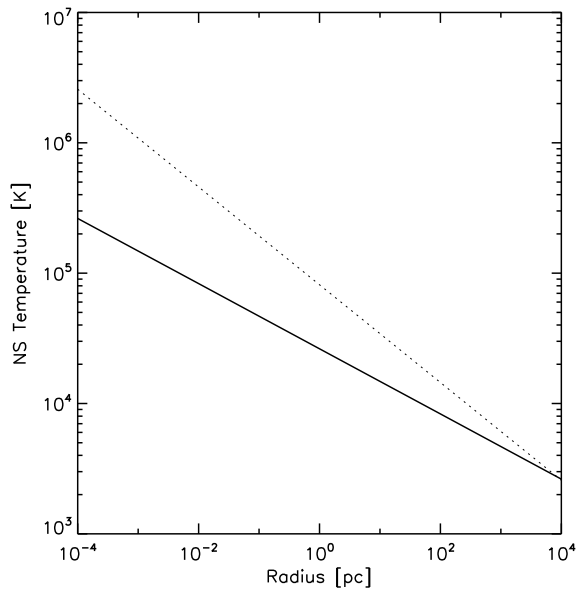} 
\caption{\it Temperature of neutron stars due to the accretion of WIMPs as a function of distance from the centre of the galaxy.  The two lines correspond to two different density profiles, $\rho\propto r^{-1}$ and $\rho \propto r^{-1.5}$.}
\label{nsheat}
\end{center}
\end{figure}

The expected heating of neutron stars is shown in figure \ref{nsheat} for two different density profiles, namely $r^{-1}$ and $r^{-1.5}$, both normalised so that the density at the solar radius 8.5 kpc from the centre of the galaxy is 0.3 GeV cm$^{-3}$.  We have assumed a cross section large enough so that the surface area of the star is the effective cross section.  We have also taken the most basic approach, namely that the neutron stars are in steady state and instantly emit the luminosity injected as black body radiation. (for a more detailed recent treatment, see \cite{kouvaris})
This figure shows that only at the solar radius can one expect the heating due to DM to be less than the minimum temperature one might expect for a neutron star which started at some high temperature 13 billion years ago (see estimate in previous section.)  It is however difficult to observe the temperature of cool neutron stars - the small radius of these objects means that only extremely hot neutron stars can be detected through their black body emission.  It might be possible to place constraints on the heating due to DM by looking at the non-observation of black body temperature from the surface of neutron stars which have been identified because they are pulsars.

The CHANDRA x-ray telescope has also observed many hot compact objects in the central parsecs of the galaxy \cite{chandra}, although any cooler neutron stars in the same region would presumably be undetectable due to the dust in the plane of the galaxy which effectively cuts out the transmission of photons below 2 keV down to energies less than the visible spectrum.  The heating effect outlined here is relatively large though and it should be possible to find some constraining situations.  

Now we will consider other effects of the build up of DM in neutron stars.  The reader should be aware of relevant previous work which considers the build up of charged DM in neutron stars \cite{nussinov}.

The timescale for DM inside the star to thermalise with the background neutrons can be estimated - DM particles falling onto a neutron star will be semi-relativistic and therefore need to lose an amount of kinetic energy roughly equivalent to their own mass through collisions with nuclei before they reach equilibrium. Kinematics show that the typical energy exchanged when the DM is more massive than nuclei will be $\Delta E\sim M_{nuc}v^2$ \cite{lewin} so that as the DM moves through the neutron matter at speed close to $c$ the timescale for it to lose energy is given by
\begin{equation}
\tau_{th}=\frac{m_{DM}}{m_{nuc}c\sigma n}
\end{equation}
where $\sigma$ is the DM nucleon cross section and $n$ is the number density of nucleons.  For example, if we would like a thermalisation time scale of less than a million years, we would require that the DM nucleon cross section $\sigma>10^{-60}(m_{dm}/m_{nuc}){\rm cm}^2$.  Once thermalised, the thermal radius within which the majority of thermalised DM particles will be located is given by
\begin{eqnarray}
r_{th}\sim \left[\frac{9kT}{4\pi G\rho_c m_{dm}}\right]^{1/2}\hspace{5cm} \nonumber\\ 
\sim 64 cm \left(\frac{T}{10^{5}\rm{K}}\right)^{1/2}\left(\frac{10^{14} \rm{g cm}^{-3}}{\rho_c}\right)^{1/2}\left(\frac{100 \rm{GeV}}{m_{dm}}\right)^{1/2}
\label{radius}
\end{eqnarray}
where we have assumed that the phase space density of DM particles is low enough that Maxwell-Boltzmann statistics are still valid, an issue we will return to below.  The number of particles required to reach equilibrium between annihilation and capture is then given by
\begin{eqnarray}
N_{eq}=5\times 10^{30}\left(\frac{\Gamma_c}{10^{29} \rm{s}^{-1 }}\right)^{1/2}
\left(\frac{T}{10^{5} \rm{K}}\right)^{3/4}\times\nonumber\\
\left(\frac{10^{15} \rm{g cm}^{-3}}{\rho_c}\right)^{3/4}
\left(\frac{100 \rm{GeV}}{m_{dm}}\right)^{3/4}\left(\frac{10^{-26} \rm{cm}^{3}\rm{s}^{-1}}{(\sigma_{ann} v)}\right)^{1/2}
\label{neq}
\end{eqnarray}
where $\sigma_{ann} v=10^{-26}$ cm$^{3}$s$^{-1}$ is the appropriate value for a thermal relic that achieve .  The capture rate of $10^{29}$ s$^{-1}$ corresponds to a rather large density of DM, but one which could be feasibly found at the galactic centre for 100 GeV WIMPs.  
Throughout the rest of this paper, we will assume an accretion rate onto the neutron star of $\Gamma_c=10^{29}(100{\rm GeV}/m_{dm})$s$^{-1}$.

\begin{figure*}
\begin{center}
\includegraphics[width=0.8\textwidth]{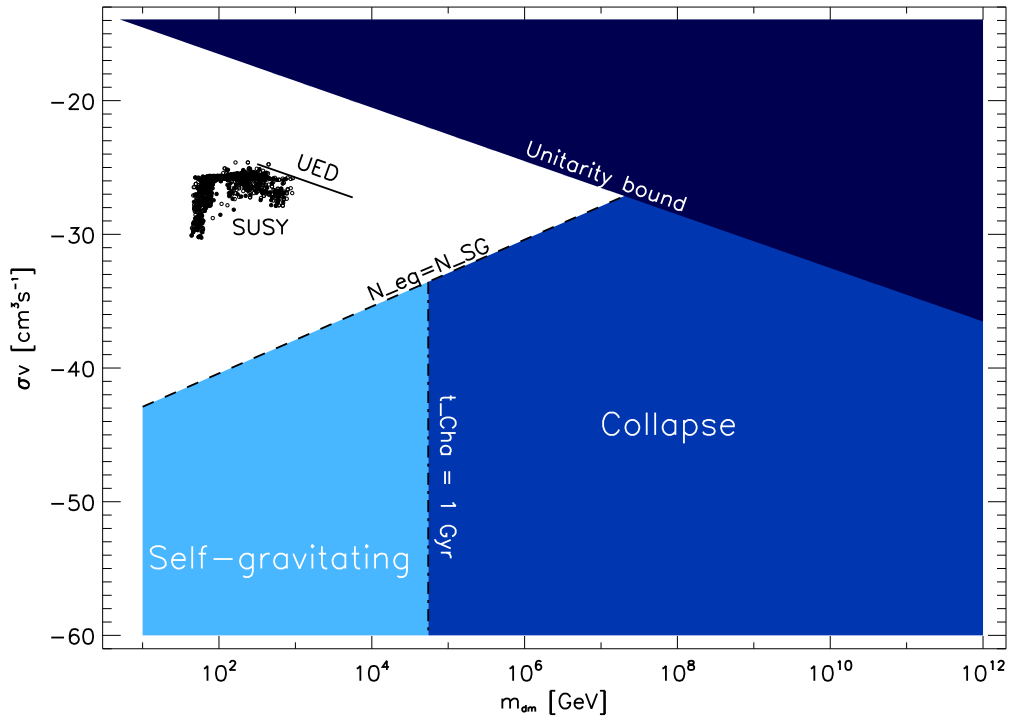} 
\caption{\label{fancy}Different outcomes of the accumulation of DM inside a neutron star, in the DM mass vs. annihilation cross section plane, for a capture rate of $\Gamma_c=({\rm 100 GeV}/m_{dm})10^{29}\rm s^{-1}$.  In the top left corner we show the results of a scan of the supersymmetric parameter space, where neutralino models are compatible with accelerator and cosmological constraints (as obtained with DarkSUSY~\cite{Gondolo:2004sc}). The solid line is relative to viable DM models in Universal Extra-dimensions (see text for further details). In the shaded region below the dashed line, DM particles become self-gravitating before equilibrium between capture and annihilation is reached. For models on the right of the dotted vertical line, particles reach the critical mass for gravitational collapse in less than 1 Gyr. These models can likely be ruled out, since they lead either to large injection of energy in the NS core, or to gravitational collapse to a Black Hole, rapidly destroying the star.}  
\end{center}
\end{figure*}

As the density of DM rises in the core of the neutron star, there will come a point at which the self-gravity of the DM core is greater than the gravity due to the baryonic matter within the same volume.  We label the number of WIMPs in the star at the moment when this occurs with $N_{SG}$, defined by $4\pi r_{th}^3\rho_c/3=N_{SG}m_{dm}$.  This means that if $N_{eq}\ge N_{SG}$ the core of DM will become self gravitating.  For a given DM particle mass, this will occur when there are
\begin{eqnarray}
N_{SG}&=&6\times 10^{40}\left(\frac{T}{10^{5} \rm{K} }\right)^{3/2}\times\nonumber\\
&&\left(\frac{10^{14} \rm{g cm}^{-3}}{\rho_c}\right)^{1/2}\left(\frac{100 \rm {GeV}}{m_{dm}}\right)^{5/2}
\label{sg}
\end{eqnarray}
DM particles in the star. The region of the parameter space where this happens in shown in fig. \ref{fancy}. 

This result only applies in cases where the Maxwell-Boltzmann statistics is relevant (assuming the DM candidates are fermions).  Consider a ball of radius $R$ made up of $N$ non-relativistic degenerate fermions of mass $m$ and temperature $T$ gravitating (not necessarily self-gravitating) due to a matter density $\rho$.  These parameters will be related by the approximate expression $R^7\rho^2 G m \sim N^{5/3}$.  The distribution of DM particles will form a degenerate 'dark star' when two conditions are fulfilled, the first is that the temperature is less than the Fermi energy, $T< n^{2/3}/m_{dm}$ which occurs for a number of DM particles greater than
\begin{equation}
N_{Fermi}=3\times 10^{37}\left(\frac{T}{10^{5} \rm{K}}\right)^{3}\left(\frac{10^{14} \rm{g cm}^{-3}}{\rho_c}\right)^{3/2}
\end{equation}
independent of the DM parameters. The second requirement is that the radius $R$ is larger than the thermal radius $r_{th}$, a situation which occurs when the number of DM particles $N$ is larger than
\begin{eqnarray}
N_{deg}&\sim& 2.5\times 10^{39}\left(\frac{T}{10^{5} \rm{K}}\right)^{21/10}\times\nonumber\\
&&\left(\frac{10^{14} \rm{g cm}^{-3}}{\rho_c}\right)^{9/10}\left(\frac{100 \rm{GeV}}{m_{dm}}\right)^{3/2} \, .
\label{degen}
\end{eqnarray}
Newly accreted DM particles would fall down onto the surface of this degenerate dark star, making it grow in mass and shrink in radius. For DM candidates above few hundred GeV, and adopting typical NS parameters, DM particles become self-gravitating before these degeneracy conditions are met. 
We also note that for a typical WIMP candidate with mass close to 100 GeV and the cosmologically favoured DM self-annihilation cross section, we do not expect the DM to form such a degenerate dark star or a self-gravitating core. However, more massive particles, especially if strongly-interacting with ordinary matter, would be severely constrained, like for example SIMPzilla particles \cite{pospelov,kolb} which are a version of superheavy DM which could be produced non-thermally during inflation and may have masses as large as $10^{12}$ GeV \cite{chung,igor}. These particles can interact with standard model matter with a cross section as large as $\sigma\sim 10^{-24}$cm$^2$ while their self annihilation cross section is limited by unitarity to be less than $(\sigma_{ann} v) = m_{SIMPzilla}^{-2}$ \cite{pospelov,kolb}. Note also that indirect detection arguments have been used recently to set an upper limit on the annihilation cross-section which is stronger than the unitarity bound for relatively light DM candidates~\cite{beacobell}. The amount of material along a path through the centre of a neutron star will correspond to $10^{44}$cm$^2$ so while each collision will only allow a maximum momentum exchange of $Q\sim m_{neutron}$ there will be enough material to slow down and stop these particles, so that the geometrical cross section will be appropriate in equation (\ref{capture}).  The capture rate in regions of the galaxy with DM densities similar to that in the solar system will therefore be $\Gamma_c\sim 10^{12}$s$^{-1}$.  Since the number of such high mass particles in a neutron star required to form a self gravitating core will be of the order of $N_{SG}=10^{16}$, such a core would quickly form in a matter of hours.  If the heavy DM candidate has a self annihilation cross section anywhere near the unitarity bound it will then instantly annihilate with itself.
A similar fate would await other heavy DM candidates although, as discussed above, DM candidates which are too light will feel their own degeneracy pressure before they collapse.  

We can estimate the maximum energy which could be released in such a burst of annihilations due to the self-gravitating collapse of a cloud of DM particles.  Equations (\ref{sg}) and (\ref{degen}) show us that the biggest release of energy in this way would be the sudden collapse and self annihilation of around $10^{36}$ DM particles of mass 10 TeV.  Such an injection of energy could in principle instigate a QCD phase transition in the central few centimetres of the neutron star, changing the equation of state in that region and possibly leading to a change in the observed rotation of pulsars.  While glitches are observed in the periodicity of pulsars \cite{glitch}, more detailed study would have to be done to quantify the magnitude and observability of any such effect. 

The equation of state of any degenerate core will of course become relativistic when the number of particles reaches the Chandrasekhar limit, which corresponds to the number
\begin{equation}
N_{Cha}\sim\left(\frac{M_{Pl}}{m_{dm}}\right)^{3}\sim 10^{51}\left(\frac{100 \rm{GeV}}{m_{dm}}\right)^{3}
\end{equation}
which for a 100 GeV DM particle is roughly to the mass of a planet like Mars.  For a SIMPzilla candidate with the properties outlined above, the Chandrasekhar limit is much less, of the order of 1000 tonnes. If that is the case, the accretion of matter onto the newly formed mini-BH would rapidly destroy the star over very short timescales (see Ref.\cite{nussinov} for a detailed discussion of the growth of the mini black-hole, and consequent destruction of the NS).

We have summarised the different behaviours in figure \ref{fancy}, where we have assumed a large DM accretion rate of $\Gamma_c=({\rm 100 GeV}/m_{dm})10^{29}\rm s^{-1}$.  If the DM densities advocated in this paper can be reached in nature, the observation of compact objects in those regions would place strong constraints on the particle physics parameters of DM particles.


\section{Conclusions}

In this work, we have investigated the observational consequences of the capture of DM onto astrophysical compact objects. Although white dwarfs and neutron stars are relatively hot, making it a challenge to detect any heating of such objects due to DM annihilations in their cores, we have presented one situation, namely the possible accretion of WIMPs onto cool white dwarf stars in globular clusters, where we argue it is feasible that such heating could be detected.  The presence or otherwise of DM in the core of globular clusters is model dependent, but we have argued that if DM does exist in the cluster, then the density at the location of the white dwarfs in question will not be a strong function of the total amount.

We have also estimated the heating of neutron stars due to the accretion of WIMP DM (see also the recent~\cite{kouvaris}). Finally, we have explored the parameter space of DM mass and self-annihilation cross section, focusing on the accretion of such particles onto neutron stars in a region of the galaxy where the density of DM is high.  We have found that there are a number of different outcomes of DM accretion, depending upon the exact parameters chosen, including the creation of degenerate dark stars inside the neutron star, or self-gravitating cores.  It would be interesting to study the fate of such configurations when they become unstable and collapse, in order to look for observable consequences.

\section*{Acknowledgments}
The authors are happy to acknowledge the support of the CERN theory group's theory institute on the LHC-cosmology interplay. We thank John Beacom, Lars Bergstr\"om, Joakim Edsj\"o, Charles Horowitz, Lam Hui and Harvey Richer for many useful discussions and comments.

\end{document}